\documentclass[pra,aps,twocolumn]{revtex4}

\usepackage{times} 
\usepackage{amsmath}
\usepackage{amssymb}
\usepackage{amsthm}
\usepackage{graphicx}
\usepackage{dcolumn}
\usepackage{bm}
\usepackage{multirow}
\usepackage{mathrsfs}


\def\unity{\mbox{\small 1} \!\! \mbox{1}}

\begin{document}
\title{Efficient growth of complex graph states via imperfect path erasure} 

\author{Earl T. Campbell}\email{earl.campbell@materials.ox.ac.uk}
\author{Joseph Fitzsimons}
\author{Simon C. Benjamin}
\author{Pieter Kok}

\affiliation{Department of Materials, Oxford University, Oxford, UK}

\begin{abstract}
Given a suitably large and well connected (complex) graph state, any quantum algorithm can be implemented purely through local measurements on the individual qubits. Measurements can also be used to create the graph state: Path erasure techniques allow one to entangle multiple qubits by determining only global properties of the qubits. Here, this powerful approach is extended by demonstrating that even imperfect path erasure can produce the required graph states with high efficiency. By characterizing the degree of error in each path erasure attempt, one can subsume the resulting imperfect entanglement into an extended graph state formalism. The subsequent growth of the improper graph state can be guided, through a series of strategic decisions, in such a way as to bound the growth of the error and eventually yield a high-fidelity graph state. As an implementation of these techniques, we develop an analytic model for atom (or atom-like) qubits in mismatched cavities, under the {\it double-heralding} entanglement procedure of Barrett and Kok [Phys. Rev. A 71, 060310 (2005)]. Compared to straightforward postselection techniques our protocol offers a dramatic improvement in growing complex high-fidelity graph states.
\end{abstract}

\maketitle

\section{Introduction}
\indent For certain algorithms, quantum computing offers the possibility of exponential speed up over classical computing if significant obstacles to physical implementation can be overcome \cite{NC01b}.  An important class of proposed implementations uses linear optical elements and photo-detection to perform the logical operations \cite{KMMRM01a}.  In these schemes the qubits are typically projected onto the required states using optical measurements. An undesired feature of this technique is that in general two-qubit gates are probabilistic.  In a naive implementation of the circuit model of quantum computation, this causes a decrease in success probability that scales exponentially in the number of two-qubit gates.  It can be avoided by dividing the circuit into sub-routines, post-selecting successful implementations of sub-routines and then teleporting the sub-routine into the main algorithm \cite{GC01a,KLM01a,YoranReznik,N01a}.  This approach is closely related to the one-way model of quantum computing developed by Raussendorf and Briegel \cite{RB01a,RBB01a, HEB01a}, and has become a serious alternative to the circuit model. 

We divide the class of optical implementations of quantum computing into purely optical schemes where the logical qubits are photons, and hybrid schemes where the logical qubits are matter systems.  In the latter class, matter systems are used to store qubits, and optical excitations followed by projective photon detections implement the two qubit gates \cite{BK01a,LBK01a,LBBKK01a,Browne,Bose,Feng,Cabrillo,BES01a,BBFM01a}.  Candidate matter qubits with optical transitions include NV centers in diamond \cite{JGPGW01a}, quantum dots in microcavities \cite{QD1} or photonic band-gap structures \cite{Photonic}, and neutral atoms in cavity QED \cite{QED, Kuhn2002}.  An argument in favour of pursuing hybrid approaches is that schemes using only linear optical elements require post-selection, and hence require a quantum memory \cite{KMMRM01a}.   The natural candidates for such memories are typically matter systems, and should therefore implement the logical qubits.  Optical delay lines are problematic as scalable quantum memories, mainly due to absorption in the fibre, and low photo-detection efficiencies.  Indeed, the problem of photon loss is a serious concern for any optical scheme, and the literature contains a variety of proposals designed to circumvent this issue \cite{RHG05,VBR06}.

\begin{figure}[t]
\begin{center}
\includegraphics{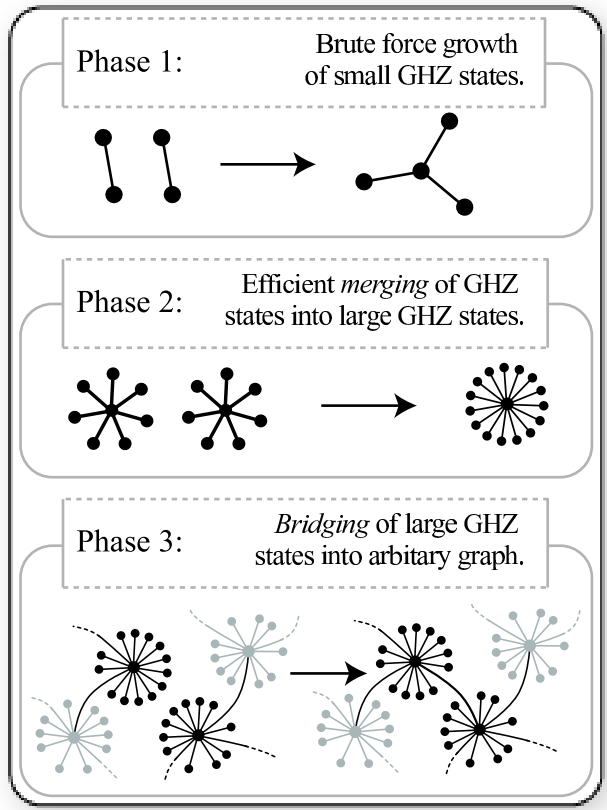}
\caption{A schematic outline of the microcluster approach to graph state growth. Phase 1: GHZ states (or microclusters) are joined into larger GHZ states, in a manner that is inefficient for building GHZ states of arbitrary size. It yields a constant offline overhead to building medium-sized GZH states.  Phase 2: Once the GHZ states are sufficiently large, they can be merged efficiently into large GHZ states. Phase 3: The large GHZ resources can be used to grow any graph (including cubic lattices like cluster states, or the minimal graph state for a particular algorithm) via a bridging procedure.}
\label{EasyAs123}
\end{center}
\end{figure}

Although hybrid schemes do not suffer from qubit loss, they still have to solve the problem of photon loss during the entangling procedures.  We can postselect successful outcomes, but photon loss can lead to a detector signature that, erroneously, indicates a successful outcome.  One hybrid scheme that achieves a degree of robustness against losing photons is the Double Heralding (DH) procedure proposed by Barrett and Kok \cite{BK01a}.  The essence of this scheme is two consecutive parity measurements that side-step any photo-collection and detection inefficiency.  In turn, this allows us to grow high-fidelity graph states for quantum computing.  In this paper, we develop upon our previous work \cite{Monitored1} to give a full model of a dominant source of errors in this scheme.  We also extend our proposed adaptive growth strategy, which can significantly alleviate the cost incurred by these errors.  The solutions proposed here are suitable for correcting any parity measurement scheme where the error is known, and they move beyond mere post-selection where efficiency is traded for fidelity. 

We address the experimental challenge that optical schemes require indistinguishable (mode-matched) photon sources.  The problem of mode matching can be divided into the matching of frequency, polarization, and spatio-temporal modes.  These categories can be further subdivided into mismatching due to random fluctuations or due to fixed variations inherent in the configuration of the matter-qubit system.  This latter category is especially important when the fabrication process for the nano-structure containing the matter qubit is not under complete experimental control (e.g., self-assembled quantum dots).  It is well known that nano-structures generally possess a very broad distribution of properties that are an intrinsic consequence of the fabrication process, but that they can be accurately characterized or calibrated.  Although here the details are worked out for temporal mismatch in the double heralding scheme, the strategies employed are more generally applicable.  Indeed, any form of mode-mismatch can be tolerated, provided that the error is a known function of some detector variable - that is, the error is \textsl{monitored}. 

\begin{center}
\begin{figure*}[t]
\includegraphics{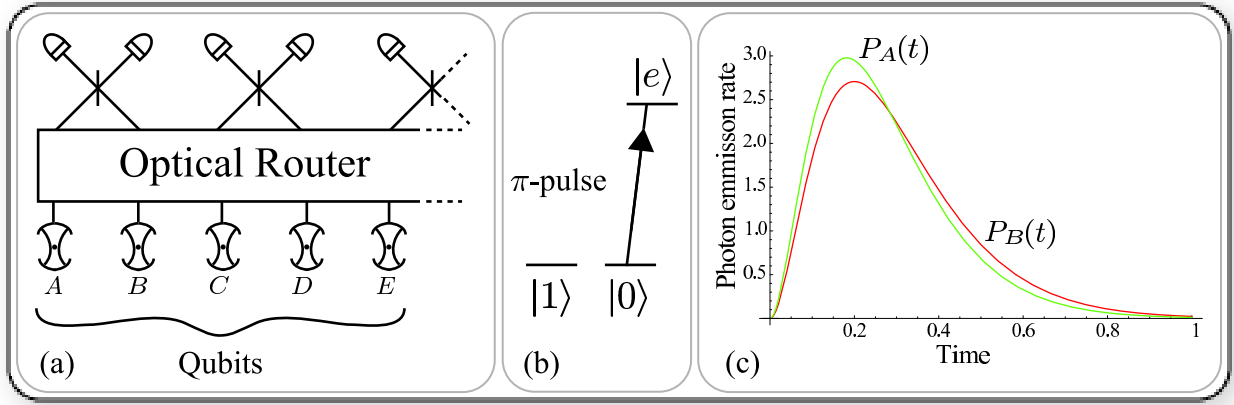}
\caption{ (\textbf{a}) A schematic layout of a typical device that could implement our proposal. Illustrated here are an array of atom/cavity systems for which optical pumping and photon measurements can be performed in parallel. A form of optical switching will also be required so that measurements can be performed on arbitrary pairs of systems. (\textbf{b}) the energy level structure for the matter qubit, which allows only the transition $\vert 0 \rangle \leftrightarrow \vert e \rangle$ via $\pi$-pulses. (\textbf{c}) Photon leakage rates $P_{A}(t)$ and $P_{B}(t)$ for two cavities $A$ and $B$.  If our scheme is not used, then to keep the errors within fault tolerance threshold of $1 - 10^{-5}$ the quantum computer must postselect on detector click times within a small time interval.}
\label{Outline}
\end{figure*}
\end{center}

More specifically, we consider the elimination of monitored errors that generate a so-called \textit{tilting} in the ideally equally weighted amplitudes of a graph state.  The error can be associated with a single graph vertex, and as such, can easily be represented in a generalized graph state notation that is introduced in section~\ref{mismatch}.  As illustrated in Fig.~\ref{Outline}a, we assume that a large array of qubits is available.  The device can perform parallel measurements between pairs of qubits, and the pairing of qubits can be optically switched.  Following Nielsen \cite{N01a}, the proposed strategy proceeds by first constructing a resource of GHZ states, or microclusters: \begin{equation}
\label{basicGHZ}
 \vert \psi  \rangle \equiv  \frac{\vert 0 \rangle^{\otimes N} + \vert 1 \rangle^{\otimes N}}{\sqrt{2}},
\end{equation} 
\noindent 
and then fusing these into an arbitrary graph.  This allows us to build a universal resource for quantum computing.  

Sections \ref{sec-BK} $\&$ \ref{mismatch} will outline the problem with a brief description of the double-heralding (DH) scheme, followed by a Jaynes-Cummings model \cite{Jaynes-Cummings} of cavity mismatch.  Sections \ref{sec-BK} $\&$ \ref{mismatch} discuss how to optimise the brute force growth of small, imperfect, GHZ states (see phase 2 in Fig.~\ref{EasyAs123}).  The error under consideration will affect these states, such that throughout phase 1, they have the more general form:
\begin{equation}
 \vert \psi (\theta_{a}) \rangle \equiv \cos (\theta) \vert 0 \rangle^{\otimes N} + \sin (\theta) \vert 1 \rangle^{\otimes N}.
\end{equation}
Here $\theta$ is the \textit{tilting} angle of the vertex, and the vertex is untilted when $\theta=\pi/4$.  Section \ref{sec:phase1GHZ} shows that when constructing a large graph with mismatched cavities, the amplitude distribution for $N$-qubit entangled qubits does not deteriorate past that of the distribution for 2-qubit construction --- although there is a modest decrease in the intrinsic gate success probability.

Section \ref{sec:Purify} shows that when these GHZ resources are large enough to be used in phase 2 of graph state growth. Those that do not meet fidelity requirements can be purified probabilistically.  We call this removal of tilting errors, \textit{realignment}.

In section \ref{ProcIntro}, two procedures are described that can fuse the purified GHZ resources.  With these procedures GHZ resources can be merged into larger GHZ states (see phase 2 in Fig.~\ref{EasyAs123}), or bridged by a graph edge, allowing construction of a general graph (see phase 3 in Fig.~\ref{EasyAs123}).  In the course of describing these procedures it will be convenient to further generalize the graphical notation to include \textit{weighted graph edges} and \textit{partial fusions}.  In section \ref{ResOverhead}, we quantify the improvements gained over a naive postselection strategy.

\section{double-heralding Scheme with Mismatched Photon Leakage Rates}
\label{secondsec}

In this section we describe the double heralding scheme with both perfectly and imperfectly matched cavities. We construct a model for cavity photon leakage and discuss how this effects the double-heralding scheme.  We discuss how the spread of the resulting tilting error can be limited, and eventually removed by the realignment procedure.

\subsection{The double-heralding scheme}
\label{sec-BK}

The double-heralding scheme for the construction of graph states \cite{BK01a, B01a} uses matter qubits that have a three-level structure denoted by the states $\vert 0 \rangle,\vert 1 \rangle$ and $ \vert e \rangle $ [see Fig.~(\ref{Outline}b)]. The low-lying $( \vert 0 \rangle , \vert 1 \rangle )$ states are the computational basis states for the logical qubit, and $ \vert e \rangle $ is an excited level that becomes occupied when a qubit in the $\vert 0 \rangle$ state is excited by a $\pi$-pulse.  Of the two computational basis states, only the $\vert 0 \rangle$ is excited by the $\pi$-pulse, as the transition from $\vert 1 \rangle$ is forbidden (e.g. by a selection rule ).

The first stage of building a graph state requires an entangling operation between two qubits.  The DH scheme for entangling two qubits consists of two rounds of measurements, and requires a resource of qubits in the $ \vert + \rangle \equiv (\vert 0 \rangle + \vert 1 \rangle)/\sqrt{2} $ state.  If one could completely prevent photon loss, and have completely reliable number-resolving detectors, then only the first step of the DH scheme would be required --- that is, the system would require only single heralding. For single heralding two matter qubits in different cavities $A$ and $B$ [see Fig.~(\ref{Outline}a)] are prepared in the $\vert + \rangle$ state and then pumped with a $\pi$-pulse.  Upon relaxation the matter qubits will emit zero, one or two photons with probabilities $25 \%$, $50 \%$ and $25\%$, respectively.  Fig.~\ref{Outline}a illustrates how a beam splitter erases the path information of the photons.  If one (and only one) photon is detected, and there is no photon loss, the system is projected onto the $\vert 0,1 \rangle \pm i\vert 1,0 \rangle$ subspace, giving a successful entangling operation. 

When photon loss or lack of photon number information is included, a single detector click will project the system onto a mixture of the ideal result and the two photon subspace, $\vert 0,0 \rangle$.  The unwanted part of the mixture can be eliminated by two more steps of the procedure.  First, both matter qubits are rotated by the Pauli spin flip matrix $X$, which does not affect the desired part of the density matrix but converts the $\vert 0,0 \rangle$ component into $\vert 1,1 \rangle$.  Next, the heralding is repeated by re-exciting the qubits with a $\pi$-pulse and waiting for any photon detection events.  If this second round of heralding produces a single detector click then the procedure has succedded; the $\vert 1,1 \rangle$ component of the mixture is eliminated, and it is known that photon loss could not have occurred on the first round.  This procedure relies on negligible dark counts in the photo-detectors \cite{BK01a}.  In order to avoid confusion, the term DH \textit{application} denotes the two \textit{rounds} of a successful entangling operation.   

After a successful application of DH the qubits are maximally entangled in the state $( \vert 01 \rangle \pm \vert 10 \rangle )/\sqrt{2} $.  Applying the local rotation $HX$ to either one of the qubits creates the graph state  $( \vert 0+ \rangle \pm \vert 1- \rangle )/\sqrt{2} $,  where $H$ is the Hadamard gate.  A larger graph can be constructed by an application of DH on two graph nodes that are both connected to only a single neighbour.  If the DH application is successful, then the resulting graph is fused \cite{B01a}.  If the DH application fails, then the two qubits must be measured in the computational basis (a Pauli $Z$ measurement) in order to remove them grom the graph state.  More generally, if the qubits have more than one neighbour each, a successful DH application will yield a graph with one qubit having all the connections of the original two qubits. The other qubit (up to a Hadamard gate) will be connected to this qubit as a single dangling bond, or a ``cherry''.

\subsection{The effect of cavity mismatch}
\label{mismatch}

The effect of mismatched cavity leakage rates on the double-heralding scheme will now be described in the Jaynes-Cummings model.  We show that mismatched cavities cause a tilting in the amplitude of the resulting state, which depends on the photon detection times.  With good time resolving photon detectors and calibrated photon sources this tilting is known to the experimenter and can be corrected by the scheme presented here.  In the non-ideal case, if atom $A$ is in a cavity that tends to emit photons faster than the cavity that hosts atom $B$, then an earlier detector click means that the photon is more likely to have come from cavity $A$, hence achieving only partial path erasure.  In order to know the resulting state, each atom-cavity system  must be calibrated by measuring the photon leakage rate from a qubit prepared in the $\vert 0 \rangle $ state. We denote this probability distribution by $P_{x}(t)$, where the $x$ indexes the atom/cavity system.  Two examples of mis-matched $P_{A}(t)$ and $P_{B}(t)$ are shown in Fig.~(\ref{Outline}c), where perfect path erasure occurs only when the curves cross.  

A suitably general model for the probability distributions $P_x(t)$ can be constructed by considering the three-level atomic system and its coupling to an electromagnetic cavity mode.  With only one quanta of energy available to the cavity mode, its quantum state will be described in the Fock basis of $\vert \varnothing \rangle$ and $\hat{a}^{\dagger} \vert \varnothing \rangle$, with no photons, and one photon of energy $\hbar \omega$, respectively.  If we monitor a joint atom-cavity system $x$ for the emission of photons, and none are detected, then its evolution can be described  by a non-Hermitian conditional Hamiltonian $\mathscr{H}_{x}$ \cite{GZ01b}:
\begin{equation}
\label{eqn:Ham}
	\mathscr{H}_{x}= g_{x}( \vert e \rangle \langle 0 \vert \hat{a} + \vert 0 \rangle \langle e \vert \hat{a}^{\dagger})  - \frac{i }{2} \hat{J}_{x}^{\dagger} \hat{J}_{x},
\end{equation}
where $\hat{J}_{x}$ is known as the quantum jump operator,
\begin{equation}	
	  \hat{J}_{x}=\sqrt{\kappa_{x}} \hat{a}\; .     
\end{equation}
The constant $g_{x}$ represents the Jaynes-Cummings coupling strength between the cavity mode and the optical transition, and $\kappa_{x}$ quantifies the leakage rate of the cavity ($\hbar=1$).  Note that the last non-Hermitian term is responsible for the irreversible evolution of the system.  This irreversible decay also reduces the norm of the wavefunction $N= \langle \Psi \vert \Psi \rangle$ at a rate $\dot{N}$ that represents the probability of detecting a photon.  In general, any measurement event associated with a jump operator $\hat{J}$ will occur with probability $\langle \Psi \vert \hat{J}^{\dagger} \hat{J} \vert \Psi \rangle$.  Defining the amplitudes of a system as follows: \begin{equation}
\label{eqn:timedep}
	\vert \Psi  \rangle_{x} = c_{1} \vert e \rangle_{x} \vert \varnothing \rangle + c_{2} \vert 0 \rangle_{x} a^{\dagger} \vert  \varnothing \rangle + c_{3}  \vert 1 \rangle_{x} \vert \varnothing \rangle,
\end{equation}
the detection probability satisfies, $\dot{N}=\kappa_{x} |c_{2}|^{2}$.  We define $P_{x}(t,g_x,\kappa_x )$ as the solution for $\dot{N}(t)$ when Eq.~(\ref{eqn:timedep}) is solved with the initial condition $c_{1}(t=0)=1$.  Note that, since $c_{1}$ is coupled to the decaying component $c_{2}$, it too will vanish over time.  

\begin{figure*}[t]
\centering
\includegraphics{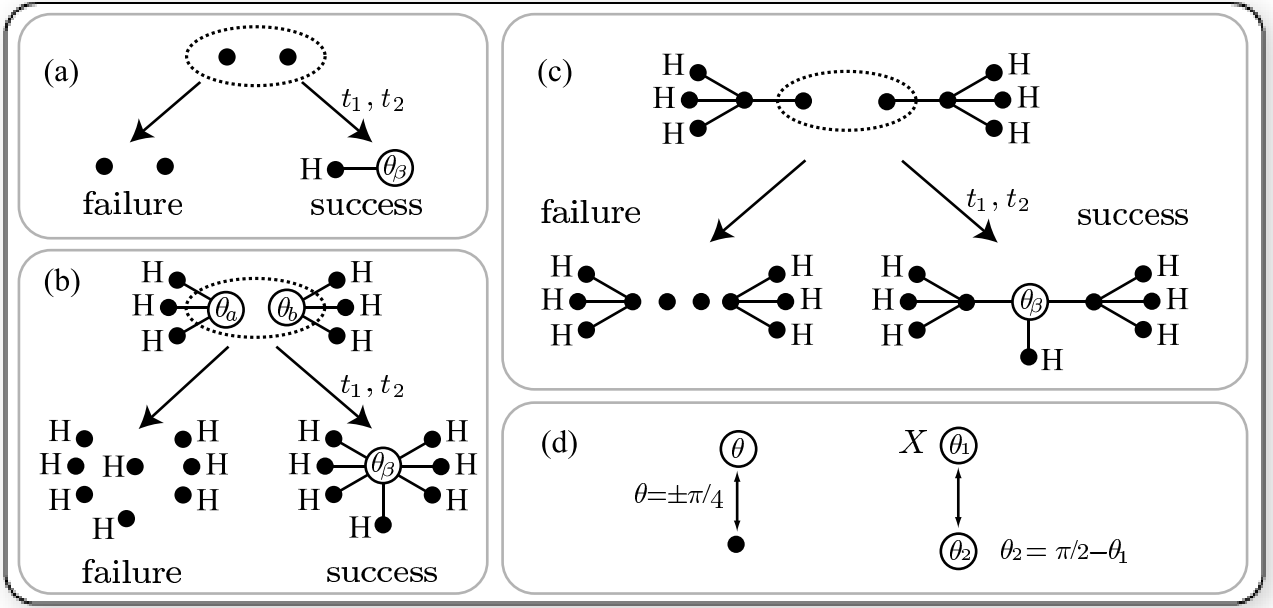}
\caption{The effect of a successful or failed application of the double heralding scheme (where the qubits used are within the dotted ellipse) for the non-ideal case: (\textbf{a}) when the initial state is two separable qubits in the untilted $\vert + \rangle$; (\textbf{b}) when the qubits are part of a tilted GHZ state; (\textbf{c}) when the qubits are part of a tilted GHZ state, but have had a Hadamard removed.  All successful outcomes have a single tilted vertex parametrised by $\theta_{\beta}$, which is specified by Eqn.~(\ref{eqn:Csqaured}).  The probability of success is given by Eqn.~(\ref{DHprob}). (\textbf{d}) some identities for tilted vertices: when $\theta=\pm\pi/4$, the graph is equivalent to a pure graph; typically in this paper we only label Hadamard rotations as they have an important effect on the double heralding scheme, although in section \ref{sec:phase1GHZ} we use the second identity illustrated --- where an $X$ rotation transforms a tilted angle $\theta \rightarrow \pi/2 - \theta$.}
\label{fig:nonIDEAL}
\end{figure*}

When a photon is detected the system is projected onto a new state by the quantum jump operator $\hat{J}_{x}$. In the case of a single photon source the constant factor $\kappa_{x}$ has no effect, but when we model the double heralding scheme it becomes physically important.

If we consider two of these systems $A$ and $B$, then whilst no photons are detected the joint system will evolve according to $	\mathscr{H}_{AB}= \mathscr{H}_{A}+  \mathscr{H}_{B}$.  If detectors $D_{+}$ and $D_{-}$ are placed behind a 50/50 beam-splitter then the corresponding jump operators $J_{+}$ and $J_{-}$ with each be a mixture of the photon modes $a$ and $b$.  To retain the same conditional Hamiltonian we require that:
\begin{equation}
\label{eqn:Jcon1}
	J^{\dagger}_{A}J_{A} + 	J^{\dagger}_{B}J_{B} = J^{\dagger}_{+}J_{+} +  J^{\dagger}_{-}J_{-} 
\end{equation}
This requirement alone does not uniquely define the new jump operators.  With the beam splitter transformations in mind, it may seem nature to presume that $J_{+}$ and $J_{-}$ will be $\sqrt{\kappa_{A}/2}(\hat{a} + \hat{b})$ and $\sqrt{\kappa_{B}/2}(\hat{a} - \hat{b})$.  However, both detectors --- being behind a 50/50 beam splitter --- have an equal chance of registering a photon, hence we require:
\begin{equation}
\label{eqn:Jcon2}
	\langle \Psi \vert J^{\dagger}_{+} J_{+} \vert \Psi \rangle = \langle \Psi \vert J^{\dagger}_{-} J_{-} \vert \Psi \rangle,
\end{equation}
which is not satisfied by the our first guess.  Jointly, the two conditions in Eqs.~(\ref{eqn:Jcon1}) and (\ref{eqn:Jcon2}) restrict the jump operators to:
\begin{equation}
	\hat{J}_{\pm} = \left( \frac{\kappa_{A}}{2} \right)^{\frac{1}{2}} a \pm \exp(i \phi) \left( \frac{\kappa_{B}}{2} \right)^{\frac{1}{2}} b 
\end{equation}
where the undetermined phase, $\exp(i \phi)$, is related to any difference in path length. This phase is unimportant as it vanishes after two rounds of double heralding. We therefore set it to 1 from now on.

Using this formalism we can now turn to modelling the effect of mismatch on the double heralding scheme.  We first consider the preparation of qubits in the state $\vert + \rangle_{A} \vert + \rangle_{B}$ and then apply the double heralding scheme.  To pass the first round the detector must click at time $t_{1}$, then be left without being interrupted by a second detector click, until the decaying amplitudes, $A_{1}, A_{2}, B_{1}, B_{2}$, are negligible.  This is will generate the state:
\begin{equation}
\vert \Psi \rangle_{AB} =  \frac{ \left( P_{A}(t_{1})^{\frac{1}{2}}\vert 0 \rangle_{A}\vert 1 \rangle_{B} + P_{B}(t_{1})^{\frac{1}{2}}\vert 1 \rangle_{A}\vert 0 \rangle_{B} \right) \vert \varnothing \rangle } {\sqrt{P_{A}(t_{1}) + P_{B}(t_{1})}}, 
\end{equation}
which lends itself to an intuitive interpretation.  Each amplitude squared is simply the relative probability of one cavity having emitted a photon compared to either cavity emitting a photon.  Proceeding with the the double heralding procedure, both qubits are flipped with $X_{A}X_{B}$, and another $\pi$-pulse is applied.  Upon detection of a second photon at time $t_{2}$, the system is projected onto:
\begin{equation}
\label{firsteq}
	\vert \Psi _f (\theta_{\beta}) \rangle_{AB} =  \cos ( \theta_{\beta} ) \vert 1 \rangle_{A} \vert 0 \rangle_{B} +
\sin(\theta_{\beta}) \vert 0 \rangle_{A} \vert 1 \rangle_{B},
\end{equation}
where the amplitudes are represented as a function of an angle, $\theta_{\beta}$, which we call the \textit{tilting} angle:
\begin{equation}
\label{tilting:simple}
	\cos (\theta_{\beta}) = \left( 1 + \frac{P_{A}(t_1) P_{B}(t_2)}{P_{B}(t_1) P_{A}(t_2)} \right)^{-\frac{1}{2}}.
\end{equation}
When $\theta_{\beta}=\pm \pi/4$, the state simplifies to a standard graph state, otherwise we say the state belongs to a class of generalized graphs that has a tilted vertex.  As shown in Fig.~(\ref{fig:nonIDEAL}a), we represent the states as graph states but with the tilted vertex labelled by an angle $\theta$.  In a similar fashion to the constructive definition of pure graph states, we constructively define the quantum state associated with a tilted graph.  Whereas a pure graph has every qubit initialised in the state $\vert + \rangle$, a tilted vertex is prepared in the state:
\begin{equation}
	\vert \theta \rangle = \cos (\theta) \vert 0 \rangle + \sin(\theta) \vert 1 \rangle\; .
\end{equation}
After this different preparation stage, control-Z gates are applied to all qubits connected by a graph edge. 

Having demonstrated the modelling techniques used on a simple two qubit system, and having defined the graphical representation of a tilted vertex, we shall now state the results for more complex systems.

Consider the effect of the double heralding scheme when the qubits used correspond to tilted vertices in tilted GHZ states.  Fig.~(\ref{fig:nonIDEAL}b) represents this in the new graphical notation (where local rotations corresponding to detector parity have been ignored).  From our constructive definition of tilted graph states it follows that the two initial states, $a$ and $b$, have the form:
\begin{equation}
\label{LopGHZ}
	\vert \psi _i (\theta_{x}) \rangle_{x} = \cos (\theta_{x})  \vert 0 \rangle^{\otimes n_{x}}_{x} + \sin(\theta_{x}) \vert 1 \rangle^{\otimes n_{x}}_{x}, \\
\end{equation}
where $x=a,b$ and $n_{a}=n_{b}=n$.  Performing a parity measurement between a qubit in $a$ with a qubit in $b$ has a success probability:
\begin{equation}
\label{DHprob}
	P(\theta_{a}, \theta_{b}) = \cos^{2}(\theta_{a})\sin^{2}(\theta_{b}) + \sin^{2}(\theta_{a})\cos^{2}(\theta_{b}),
\end{equation}
\noindent which is the upper bound of success for the double heralding scheme --- reduced upon consideration of photon loss.  Upon failure the two qubits are measured in the $Z$ basis, projecting all $2n$ qubits into a separable state.  Success yields a single $2n$ qubit entangled state:
\begin{eqnarray}
\label{eqn:BKeprPAIR}
	\vert \Psi _f(\theta_{\beta}) \rangle_{xy} & = & \cos(\theta_{\beta})  \vert 0 \rangle^{\otimes n}_{x} \vert 1 \rangle^{\otimes n}_{y} \\ \nonumber
	& & + \sin(\theta_{\beta}) \vert 1 \rangle^{\otimes n}_{x} \vert 0 \rangle^{\otimes n}_{y} ,
\end{eqnarray}
which is, up to a spin flips on $n$ qubits, equivalent to another tilted GHZ state with a tilted vertex parametrised by $\theta_{\beta}$, such that:

\begin{equation}
\label{eqn:Csqaured}
	\cos(\theta_{\beta}) =  \left( 1 + \frac{\tan^2(\theta_{b})P_{A}(t_1) P_{B}(t_2)}{\tan^2 (\theta_{a}) P_{B}(t_1) P_{A}(t_2)} \right)^{-\frac{1}{2}} . 
\end{equation} 

This can be shown to be consistent with the less general result when the initial states are pure graph states, by substituting in $\theta_{a}=\theta_{b}=\pi/4$ and deriving (\ref{tilting:simple}).  Furthermore this simplification is possible when $\theta_{a}=\theta_{b}$, which is a point that we shall return to in section (\ref{sec:phase1GHZ}).

In Fig.~(\ref{fig:nonIDEAL}.c), the graphical notation is used to describe how the double heralding scheme applies when the qubits used are not the nodes of a GHZ state.  Note that the notion of central node requires some clarification, since the state described by Eqn~(\ref{LopGHZ}) has no single qubit playing a privileged role that would make it `the' node.  Indeed, applying the double heralding scheme will have the same effect independently of which qubit is used.  However, in Fig.~(\ref{fig:nonIDEAL}.c) the qubits used possess an additional Hadamard rotation compared to Eqn~(\ref{LopGHZ}).  This has an effect on the shape of the graph for successful and failed outcomes.  However, it has no effect on the probability of success, or the value of the resulting tilted vertex, which both still obey Eqs.~(\ref{DHprob}) and (\ref{eqn:Csqaured}).

\begin{figure*}[t]
\centering
\includegraphics{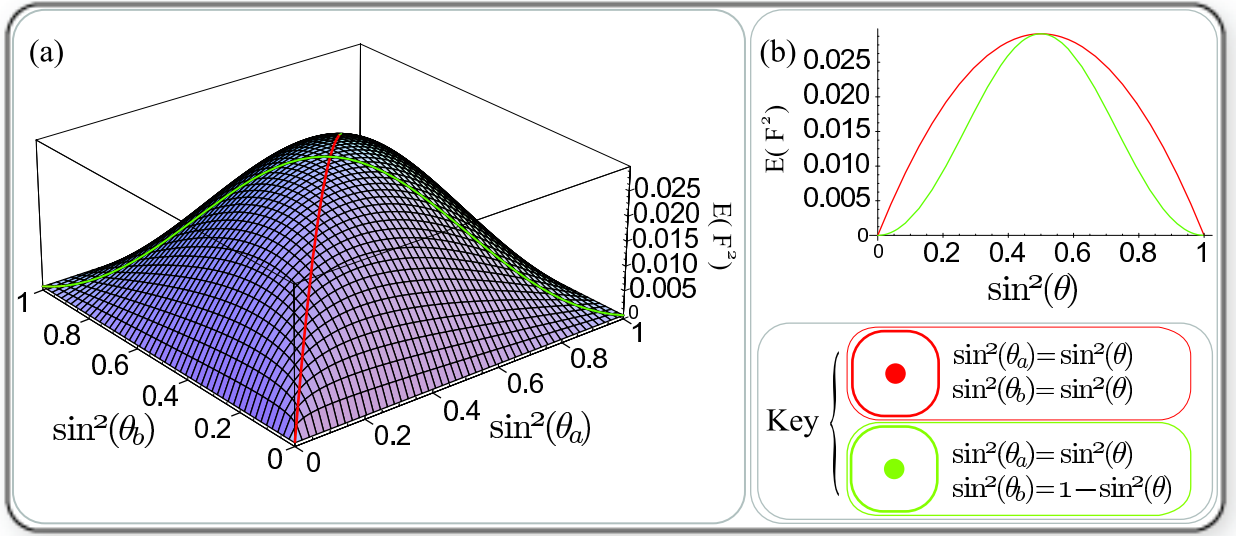}
\caption{Plots of $E(F^{2})$, the expectation value for the fidelity squared. The example cavities used are described in appendix.~(\ref{sec:CavityExample}). (\textbf{a}) A 3D-plot of the solution for $E(F^{2})$ as a function of $\sin^{2}(\theta_{a})$ and $\sin^{2}(\theta_{b})$. The red and green lines correspond to the cross sections used in (b). (\textbf{b}) Two cross sections of the 3D-plot, along $\sin^{2}(\theta_{a})=\sin^{2}(\theta_{b})$ (red) and $\sin^{2}(\theta_{a})=1-\sin^{2}(\theta_{b})$ (green).}
\label{fig:FidelitySQ}
\end{figure*}

So far we have given a full account of the effect of cavity mismatch when the detectors click at times $t_{1}$ and $t_{2}$. What remains to complete the description is to specify the probability distribution for the random variables $t_{1}$, $t_{2}$.  The probability density for a single detector click in round one is:
\begin{eqnarray}
\label{eqn:P1}
	 Q_{1}(t_1) & = & \cos^{2}(\theta_{a}) \sin^{2}(\theta_{b})P_{A}(t_1) \\
	 		& + & \sin^2(\theta_{a}) \cos^{2}(\theta_{b})P_{B}(t_1). \nonumber
\end{eqnarray}
Given an event at $t_{1}$, the second round is governed by the probability density distribution:
\begin{equation}
\label{eqn:P2}
	Q_{2}(t_2 | t_1)  =  \frac{(X + Y)P_{A}(t_{2})P_{B}(t_{2})} {X P_{A}(t_{2}) + Y P_{B}(t_{2})}.
\end{equation}
where
\begin{eqnarray}
\label{eqn:XandY}
	X & = & \cos^{2}(\theta_{a}) \sin^2(\theta_{b})  P_{A}(t_{1})  P_{B}(t_{2}) ~\text{and} \\
 	Y	& = & \sin^{2}(\theta_{a}) \cos^2(\theta_{b})  P_{B}(t_{1})  P_{A}(t_{2}), 
\end{eqnarray}
These are convenient variables that will allow us to cast many expressions in a concise form.  The product of these distributions gives the joint probability density distribution for $t_{1}$ and $t_{2}$:
\begin{equation}
	Q_{12}(t_{1}, t_{2}) =  Q_{1}(t_{1}).Q_{2}(t_{2}|t_{1})	= X + Y\; .
\end{equation}

\subsection{Expectation values for some measures of gate quality}
\label{sec:measures}

In this section we consider the expected quality of an attempt at double heralding.  The results of this section will form the criteria for evaluating different strategies for graph growth that are discussed in the subsequent section.  We consider two tilted vertices $\theta_{A}$ and $\theta_{B}$, with two leakage rates $P_{A}(t)$ and $P_{B}(t)$, from which we can determine the probability distribution for $\theta_{\beta}$.  Our quality measure must incorporate both the probability of success, defined in Eqn.~(\ref{DHprob}), and some function that measures how close $\theta_{\beta}$ is to the ideal values of $\pm \pi/4$. Without loss of generality, we shall only consider $+ \pi/4$ as the ideal case, and assume known phase errors are always corrected with a local $Z$ rotation.

Although we could use a pure state entanglement measure, such as the von-Neuman entropy, we shall instead consider functions of the inner product between the tilted state and the ideal state:
\begin{eqnarray}
	f(\theta_{\beta}) & = & | \langle \Psi (\theta_{\beta}) \vert \Psi (\pi/4) \rangle  |^{2} \\ \nonumber
										& = & \frac{1}{2}(1 + \sin(2 \theta_{\beta})).
\end{eqnarray}
In this section we will give some plausibility arguments for the value of these measures. However they are ultimately established by the nature of the realignment procedure, which distils the tilted states and is the topic of section.~\ref{sec:Purify}.

Since even a failed double heralding application generates $\vert \Psi(0) \rangle$, which has $f(0)=1/2$, we choose the quantity of interest to be $F(\theta_{\beta})=f(\theta_{\beta})-1/2$.  Using Eqn.~(\ref{eqn:Csqaured}), $\theta_{\beta}$ is eliminated from $F(\theta_{\beta})$ to give:
\begin{equation}
	F =\frac{\sqrt{XY}} {X + Y}  ,
\end{equation}
where $X$ and $Y$ have been defined in Eqn.~(\ref{eqn:XandY}).  To get the average fidelity over all possible detector click times for $t_{1}$ and $t_{2}$, we multiply by $Q_{12}(t_{1}, t_{2})$, which takes the simple form $Q_{12}=X + Y$, and integrate over $t_{1}$ and $t_{2}$:
\begin{equation}
	E ( F ) = \frac{1}{4} \sin ( 2 \theta_{a}) \sin( 2\theta_{b} )\left( {\int \left[ P_{A}(t)P_{B}(t) \right]^{\frac{1}{2}} dt}, \right)^{2}\; .
\end{equation}
Hence the expected fidelity can be split into two independent factors, one dependent on initial tilting angles, and one dependent on the overlap between the probability distributions.

However, many different fidelity distributions may have the same expected fidelity and yet represent very different resources.  Consider two distributions: the set $U$ of 2 n-qubit tilted graphs all with $\sin(2 \theta_{U}) = A/2$; and the set $V$ of 4 n-qubit tilted vertices with $\sin(2 \theta_{V}) = A/4$.  If we sum the values of $F$ they both equal $A$, but the two distributions are of different utility in subsequent attempts at double heralding.  Attempting to double herald set $U$, will produce a distribution of 2n-qubit tilted graphs with expected fidelity proportional to $\sin^{2}(2 \theta_{U}) = A^{2}/4$. Whereas the same calculation for set $V$, gives $2 \sin^{2}(2 \theta_{V}) = A^{2}/8$.  Heuristically, this indicates that quality is better than quantity.  Quantitatively, it tells us that $F^{2}$ may be a better measure of gate quality.  As alluded to earlier, this is a point that is supported by section.~\ref{sec:Purify}.  The expected value of $F^{2}$ is:
\begin{equation} 
	E ( F^{2} )  = \frac{1}{4} \int \frac { \Theta_{1} 	\Theta_{2} P_{A}(t_{1})P_{B}(t_{2})P_{B}(t_{1})P_{A}(t_{2}) dt_{1} dt_{2}}{\Theta_{1} P_{A}(t_{1})P_{B}(t_{2}) + \Theta_{2}	P_{B}(t_{1})P_{A}(t_{2}) }  ,
\end{equation}
where,
\begin{eqnarray}
\label{eqn:BigTheta}
	\Theta_{1} & = & \cos^{2}(\theta_{a}) \sin^{2} (\theta_{b}), \\ \nonumber
	\Theta_{2} & = & \sin^{2}(\theta_{a}) \cos^{2} (\theta_{b}). 
\end{eqnarray}
Unfortunately, the general case does not admit tilting angles to be factored outside the integral.  Furthermore, even for specified cavities, evaluating the integral can be quite involved.  We proceed by considering an example, for a particular pair of cavities.  The details of these cavities are given in appendix~\ref{sec:CavityExample}, and are the same as used to generate the example probability distributions given in Fig.~(\ref{Outline}).  In appendix ~\ref{Integral}, we demonstrate the details of how to evaluate $E(F^{2})$, and we give a general proof that for any probability distribution the behaviour of $E(F^{2})$ is qualitatively the same.  Hence, we can draw conclusions from Fig.~(\ref{fig:FidelitySQ}) without concern that they do not carry across for other pairs of cavities.  

\subsection{Strategies for Phase one GHZ growth}
\label{sec:phase1GHZ}

This section proposes strategies for the ``phase 1'' growth of GHZ states on the basis of $E(F^{2})$ as a measure of the expected gate quality.  Firstly, we consider the procedure given two tilted GHZ states with tilting angles $(\theta_{a}, \theta_{b})$ and cavity leakage rates $P_{A}(t)$ and $P_{B}(t)$.  Secondly, we discuss the more complex issue of how best to divide $2N$ GHZ states from a set $(\theta_{1},....\theta_{2N})$ into $N$ pairs of GHZ states  --- where the objective is that the selection maximises $E(F^{2})$ summed over all pairs. 

Consider the first problem, for which the only freedom we have that can change the outcome is the application of local rotations prior to a double heralding application.  If we apply an $X$ rotation to one of the qubits, then up to some local rotations on other qubits in the graph this interchanges the magnitude of the qubits $\vert 0 \rangle$ and $\vert 1 \rangle$ --- the tilting angle transforms $\theta \rightarrow \pi/2 - \theta$.  Hence, if $\theta_{a}$ and $\theta_{s}$ are of similar magnitude, they become dissimilar, and vice-verse.  This also inverts the success and failure probabilities, so applying spin flips alternates between high and low success probabilities.  When the success probability is high, one qubit is much more likely to emit a photon than the other, and we can expect the resulting graph to be more tilted.  Interestingly, the $X$ flips have no effect on $E(F)$, as these two effects cancel out exactly.  However, $E(F^{2})$ is more sensitive to fidelity than success probability.  Guided by this measure, we prescribe that a spin flip is applied when the tilting angle are initially far apart, $|\sin^{2}(\theta_{a})- \sin^{2}(\theta_{b})|>1/2$.  To put this argument on a more quantitative grounds, we consider when the tilting angles are symmetric about $\pi/4$, and hence far apart.  Now $E(F^{2})$ is the green curve of Fig.~(\ref{fig:FidelitySQ}), and an $X$ flip changes makes the tilting angles identical and changes $E(F^{2})$ to the red curve.

\begin{figure*}[t]
\centering
\includegraphics{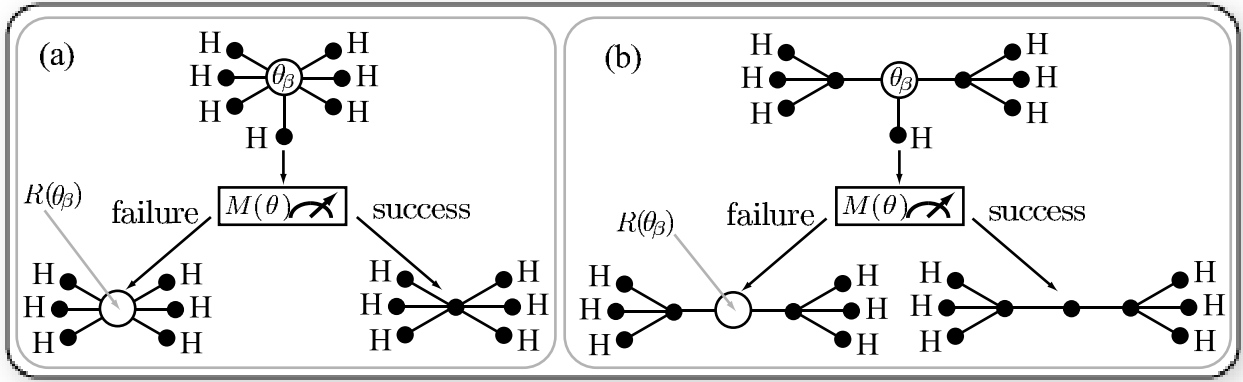}
\caption{The realignment procedure applied to: (\textbf{a}) any qubit in a GHZ state, as all qubits in a GHZ state are cherries; (\textbf{b}) the cherry of the inter-node tilted vertex.  The rotation required before a computational basis measurement is $M(\theta_{\beta})$, and this is defined by Eqn.~(\ref{HA}).  The procedure secedes with probability $p_{s}(\beta)$, Eqn.~(\ref{eqn:Ps}).  Upon failure the tilting in exacerbated, such that the vertices is tilted by angle $R(\theta_{\beta})$, Eqn.~(\ref{eqn:FailureFunction}).}
\label{fig:Realign}
\end{figure*}

We now turn to the second question: how best to pair up a set of tilted GHZ states ($\theta_{a}, \theta_{b},... \theta_{2N}$) for double heralding.  The optimal strategy is computationally hard because it requires evaluating a complex measure of success, like $E(F^{2})$, for every possible combination of pairs.  Furthermore, since the quantum computer will have to perform these assessments whilst running, time is a critical factor, as delays between rounds of double heralding will increase the amount of decoherence.  Non-optimal, yet good strategies that are computationally efficient to implement may be divided into two categories: those based solely on matching cavities; and those based solely on tilting vertices.  It may be the case that there also exist computationally efficient strategies that use both pieces of information, although whether this is the case is not clear at this point.

In this paper we propose a strategy that takes only tilting angles into consideration.  We have chosen this over a cavity based strategy for two reasons: (1) Although we may match cavities during GHZ growth, to make a unified graph we will eventually have to connect qubits from mis-matched cavities; (2) The problem does not admit a general answer, as it depends on the connectivity of the target graph, and the distribution of cavities used.

Having motivated our proposed strategy, its enunciation is simple.  The tilted GHZ states are sorted into an ordered list from descending to ascending tilting angle, and the adjacent items on the list are paired up.  Again this utilizes the fact that the red curve in Fig.~(\ref{fig:FidelitySQ}) is higher than the green curve, and this strategy will sort the qubits to be as close the the red line as possible.  

As the size of the quantum computer grows, so too will the number of entangling operations being carried out in parallel.  Applying the proposed strategy in this limit of many parallel operations approaches a situation where all vertices are paired with other vertices of an identical tilting angle.  In this limit the expression for the tilting angle after double heralding  (Eqn.~(\ref{eqn:Csqaured})) reduces to the result for when pure, untilted, graph are used.  Therefore, the average fidelity of GHZ states will not deteriorate past the fidelity distribution of building 2-qubit tilted graphs.  However, the scheme does not mask the reduced probability of success.  

\subsection{The realignment procedure}
\label{sec:Purify}

After a supply of entangled qubits of the required size has been produced, those GHZ states that do not meet a criterion of acceptable fidelity can be purified by a probabilistic procedure described in this section. A graphical description of this procedure is shown in Fig.~(\ref{fig:Realign}).  One qubit of each low fidelity GHZ state is rotated by the unitary matrix $M(\theta)$: 
\begin{equation}
\label{HA}
M(\theta) = \left( \begin{array}{cc}
- \cos (\theta) & \sin (\theta) \\
\sin (\theta) & \cos (\theta) \end{array} \right)\ .
\end{equation}
For realignment the variable $\theta$ is set equal to $\theta_{a}$, where for the rotated qubit $\cos(\theta_{a})$ is the amplitude of the $\vert 0 \rangle$ component.  Note that in the ideal limit, as $\theta_{a} \rightarrow - \pi/4$, $M(\theta_{a})$ becomes the Hadamard gate.  Following this rotation the qubit is measured in the computational basis.  With probability $p_{s}(\theta_{a})$, where:
\begin{equation}
\label{eqn:Ps}
	p_{s}(\theta)= \frac{1}{2} \sin^{2} (2 \theta),
\end{equation}
the state $\vert 1 \rangle$ is measured, and the remaining qubits are projected into a maximally entangled GHZ state.  Notice that the probability of success is proportional to $F^{2}$.  The expectation value $E(F^{2})$ was proposed in section \ref{sec:measures} as a sensible measure of how useful we can expect the product of a given double heralding application.  It should now be clear that $F^{2}$ is a good measure, since it tells us how many untilted GHZ states we can expect if we attempt to double herald and then attempt to realign.

If the procedure fails then the amount of lopsidedness of the GHZ state increases, such that the tilting angle changes to $-R(\theta_{a})$, where the function $R(\phi)$ is defined such that:
\begin{equation}
\label{eqn:FailureFunction}
		\cos \left( R (\phi) \right) = \frac{\cos^{2}(\phi)}{\sqrt{1-\frac{1}{2}\sin^{2}(2 \phi )}}\; .
\end{equation}

The success probability has an upper bound of $1/2$, which is approached as $\cos^{2}(\theta_{a}) \rightarrow 1/2$.  On first inspection this contradicts the fact that in the ideal limit an $X$ basis measurement will deterministically remove one of the qubits to give another GHZ state.  However, when $\vert 0 \rangle$ is measured on a state for which $\cos^{2}(\theta_{a}) = 0,1/2,1$, the amount of entanglement lost drops to zero (since, $-R(\theta_{\alpha})=-\theta_{a}$, which is the same state up to a local $Z$ rotation).  If the first attempt at realignment fails, a single qubit is lost but the procedure can be reattempted on the remaining qubits with a lower success probability corresponding to the new tilting angle.

In addition to purifying tilted GHZ states, the realignment procedure is applicable to a wide class of tilted graph states.  A second example is given in Fig.~(\ref{fig:Realign}b), which begins with the tilted graph that was generated as the failed outcome of the procedure shown in Fig.~(\ref{fig:nonIDEAL}c).  The tilted vertex has a single neighbour, which itself has no other neighbours, which we refer to as a \textit{cherry}.  It is the cherry that is measured out in the realignment procedure, and in general, this the is graph resource required to attempt to realign a tilted vertex.  

As a closing remark it is worth noting that the rotation $M(\theta)$ and its relationship to $R(\theta)$ go beyond the realignment procedure.  Indeed, the $M(\theta)$ rotation plays a pivotal role in both the merge and bridge procedures of later section.  Furthermore, the function $R(\theta)$ will appear in the state description of failed procedures.

\section{Procedures for joining resource states}
\label{ProcIntro}

\begin{figure}[t]
\centering
\includegraphics{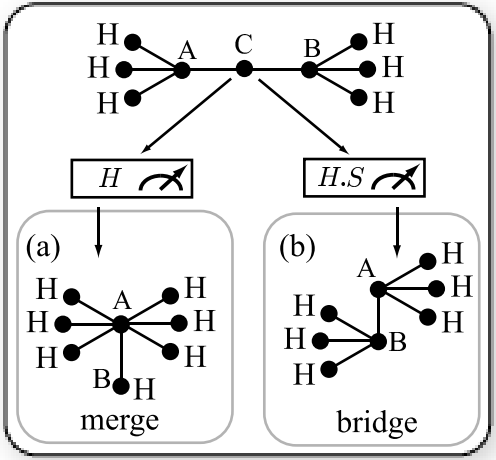}
\caption{The starting graph of this figure is the result of a successful double-heralding application to two qubits, with Hadamards, from GHZ states (with one of the inter-node qubits removed by an $X$-basis measurement).  The insets indicate how from this graph we can deterministically \textit{merge} or \textit{bridge}, $A$ and $B$. (\textbf{a}) To merge: a $X$-basis measurement is performed on the qubit C; (\textbf{b}) To bridge: a $Y$-basis measurement is performed on qubit C.}
\label{Procedures}
\end{figure}

So far three things have been demonstrated: (\textit{i}) during phase 1: the degree of tilting in the growth of small GHZ states can be limited; (\textit{ii}) at the end of phase 1: tilted GHZ states can be realigned into pure graph states (\textit{iii}) when joining GHZ states, (for phase 2 or 3), there is one available cherry with which to attempt realignment.  The remainder of this paper concerns what happens after (\textit{iii}), that is, what can be done with the tilted graph state after the failed realignment of Fig.~(\ref{fig:Realign}b).  The naive, and inefficient, answer is that the central qubit can be measured out and we can try again.  However, in this section we propose two procedures that probabilistically utilise this tilted vertex.  The procedures are coined the \textit{merge} and \textit{bridge} procedure which, when successful, fuse the nodes to which the central qubit was connected.  Which procedure should be used depends on the target graph as both generate different kinds of fusion.  The effect of a successful procedure is best seen by comparison with the effect of an $X$ or $Y$ measurement on an ideal graph, as shown in Fig.~(\ref{Procedures}).  A successful bridge procedure will, like a Y-basis measurement, connect the nodes by a graph edge.  A successful merge procedure will, like an X-basis measurement, redundantly encode the two nodes in the same way as type-II fusion \cite{BR02a}.

As a consequence of a failed attempt at either of these procedures the resulting state is not a pure graph state.  However, the failure outcomes can still be described by a further extension to our graphical language where failed mergers and bridges create \textit{partial fusions} and \textit{weighted graph edges}, respectively.  In later sections, we show that even these improper graph states are of use, as the partial entanglement established by a partial fusion or weighted graph edge can be recycled to improve the success probability of subsequent attempts at merging or bridging. This extended graphical language will be introduced with each procedure.  

\begin{figure*}[t]
\centering
\includegraphics{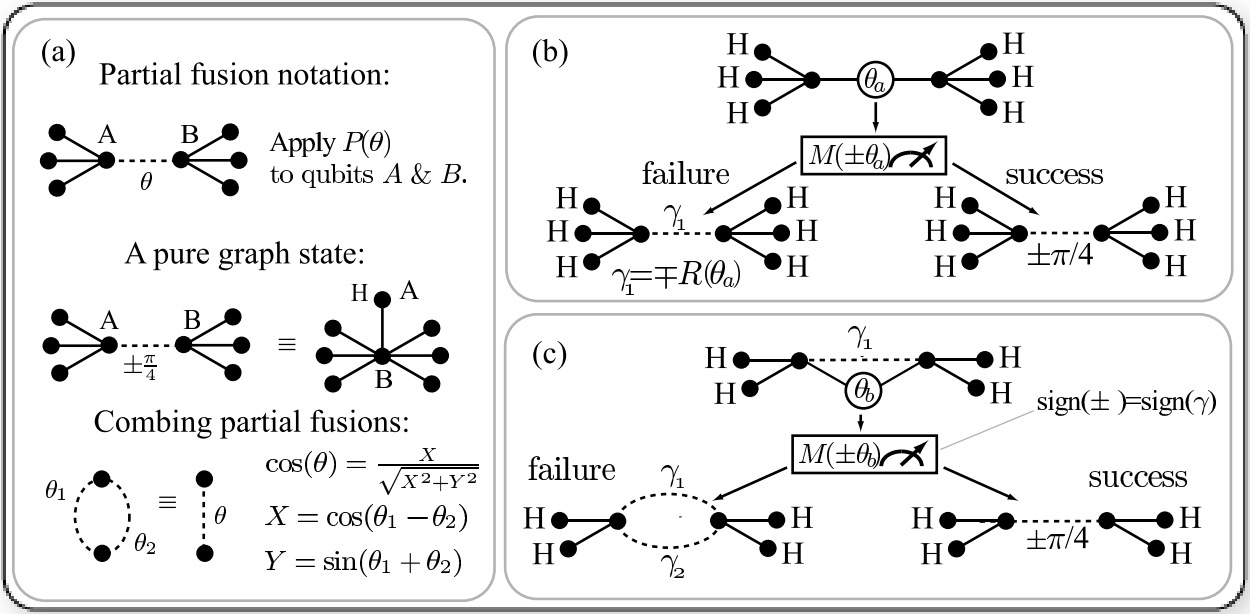}
\caption{The graphical notation for a partial fusion is defined and the procedure for merging tilted vertices is shown.  (\textbf{a}) The symbol for a partial fusion, a dashed graph edge labelled with an angle.  We also show how a maximally entangling partial fusion ($\theta=\pm \pi/4$) reduces to a pure graph state, and how multiple partial fusions combine into a single partial fusion. (\textbf{b}) The procedure for merging tilted vertices when there is no pre-existing partial fusion. (\textbf{c}) The procedure for merging tilted vertices when there is a pre-existing partial fusion.  The rotation required before measuring the tilted vertex is the same in both (b) and (c); $M(\pm \theta)$, as in Eqn.~(\ref{HA}). Upon failure a partial fusion is generated of angle $\mp R(\theta)$, where $R(\theta)$ is the failure function of Eqn.~(\ref{eqn:FailureFunction}). Note that, to maximise the probability of success the sign choice in the rotation $M(\pm \pi/4)$ must match the sign of the pre-existing partial fusion.}
\label{fig:merge}
\end{figure*}

\subsection{Merging, and partial fusions}
\label{sec:merge1}

This section describes how the merge procedures works, and how both outcomes are represented as partial fusions.  We begin with the system corresponding to the failure outcome of Fig.~(\ref{fig:Realign}b):
\begin{equation}
\label{eqn:imperfectfuse}
	\vert \Psi \rangle = \left( \cos ( \theta_{a} ) \vert 0 \rangle + \sin (\theta_{a})\vert 1 \rangle  Z_{x}Z_{y} \right) \vert \psi_{X} \rangle \vert \psi_{Y} \rangle ,
\end{equation}
where the first qubit is the tilted central vertex that has $x$ and $y$ as neighbours. These are themselves part of two other graph segments $\vert \psi_{X} \rangle$ and $\vert \psi_{Y} \rangle$.  In the case of Fig.~(\ref{fig:Realign}b), the graph segments have the form:
\begin{equation}
	\vert \psi_{X} \rangle=  \frac{1}{\sqrt{2}} \left( \vert 0^{\otimes 4} \rangle  +  \vert 1^{\otimes 4} \rangle \right)\; .\\
\end{equation}
The merging procedure requires only that $\vert \Psi_{X} \rangle$ and $\vert \Psi_{Y} \rangle$ are graph states that contain qubits $x$ and $y$.

The merger procedure is represented in Fig.~(\ref{fig:merge}b) and begins with a rotation of the central qubit by $M(\pm \theta_{a})$, where $\theta_{a}$ parametrises the tilting of the central vertex.  This gives the state:
\begin{eqnarray}
\label{Xprob}
	\vert \Psi \rangle & = &  \sqrt{p_{s}(\theta_{a})}\vert 1 \rangle P_{xy}(\pm \pi/4) \vert \psi_{X} \rangle \vert \psi_{Y} \rangle \\ \nonumber
	& & - \sqrt{1-p_{s}(\theta_{a})} \vert 0 \rangle  P_{xy}(\gamma) \vert \psi_{X} \rangle \vert \psi_{Y} \rangle \; ,
\end{eqnarray}
where $\gamma$ is related to the failure function introduced in the realignment section, $\gamma=\mp R(\theta_{a})$, and the operator $P_{xy}$ is given by
\begin{equation}
\label{P0}
P_{xy}(\phi)= \cos(\phi)\unity + \sin(\phi)Z_{x}Z_{y}\; .
\end{equation}
For $\phi=\pm \pi/4$ this reduces to the even (+) or odd (--) parity projector,
\begin{equation}
\label{P1}
P_{xy}(\pm \pi/4) = \frac{\unity \pm Z_{x}Z_{y}}{\sqrt{2}}\; .
\end{equation}
Note that we are using the expression parity projector loosely, as $P^{2}_{xy}(\pm \pi/4)$ is only equal to $P_{xy}(\pm \pi/4)$ up to a constant.  The reason for the unconventional normalization of $P_{xy}(\theta)$ is so that when it acts on the graph state $\vert \Psi_{X} \rangle \vert \Psi_{Y} \rangle$, it generates a normalized state, hence the probability of success can be read of as the square of the amplitude.

A measurement in the $Z$-basis yields a successful merge when the $\vert 1 \rangle$ state is measured, which happens with probability $p_{s}(\theta)$.  The result is that $x$ and $y$ are projected by an even (+) or odd (--) parity projector.  Since the choice of sign in $M(\pm \theta)$ determines the parity subspace of a successful outcome, we say that $M(\pm \theta)$ targets a particular subspace.  In the ideal limit where $\theta \rightarrow \pi/4$, the success probability $p_{s}(\theta)\rightarrow 1/2$. However, we should expect it to tend to 1 as the ideal case is deterministic.  This paradox is resolved when we notice that, as $\theta \rightarrow \pi/4$, the failure outcome $P_{xy}(\gamma)$ tends towards the non-targeted parity projector $P_{xy}(\mp \pi/4)$, and hence even failure becomes a success.  Indeed, the merging procedure is completely continuous with an $X$-basis measurement, as the prescribed rotation becomes the Hadamard under ideal conditions.  

Fig.~(\ref{fig:merge}) casts this procedure into a graphical language by defining a new kind of graph edge, represented by a dashed line, that is labelled by an angle $\theta$.  We call this dashed line a \textit{partial fusion}, which is to be interpreted as meaning that qubits it connects have the operation $P_{xy}(\theta)$ applied to them.  Furthermore, since partial fusions with $\theta=\pm \pi/4$ generate pure graph states, they are called pure fusions.  To see how a pure graph state can regained after an impure fusion has occurred, it is first necessary to see how partial fusions combine.  Consider two partial fusions $P_{xy}(\theta_{1})$ and $P_{xy}(\theta_{2})$, which are acting on the same qubits as in Fig.~(\ref{fig:merge}a), then:
\begin{equation}  
  	P_{xy}(\phi_{1})P_{xy}(\phi_{2}) = \cos(\phi_{1}-\phi_{2}) \unity + \sin (\phi_{1} + \phi_{2}) Z_{x}Z_{y},
\end{equation}
which can be expressed as a new partial fusion ( plus a renormalization constant ):
\begin{equation}
\label{comPF1}
	 	P_{xy}(\phi_{1})P_{xy}(\phi_{2}) =  N_{M}(\phi_{1},\phi_{2}) P_{xy}(\phi),
\end{equation}
where,
\begin{equation}
\label{comPF2}
		\sin(\phi) =  \sin ( \phi_{1} + \phi_{2}) / N_{M}(\phi_{1},\phi_{2}) ,
\end{equation}
and,
\begin{equation}
\label{comPF3}
		N^{2}_{M}(\phi_{1},\phi_{2}) =  \cos^2 (\phi_{1}- \phi_{2}) + \sin^2 (\phi_{1} + \phi_{2}).
\end{equation}
One significance of these relations is that a pure partial fusion always overrides a partial fusion, $P_{xy}(\pm \pi/4) P_{xy}(\phi) = P_{xy} ( \pm \pi/4)$, as would be expected.  Furthermore, the behaviour of the normalisation constant is crucial since the normalisation constant determines the probability of success.  The nature of this dependence will be made explicit in the following disscussion.

We now consider what can be made of the improper graph state which forms the failure outcome of Fig.~(\ref{fig:merge}b), and the procedure that we shall derive is show in Fig.~(\ref{fig:merge}c).  In preparation, another attempt at double heralding has to be made on two qubits from $\vert \Psi_{X} \rangle$ and $\vert \Psi_{Y} \rangle$. When successful, this produces another tilted central vertex connected to qubits $x$ and $y$.  The state of this system is now:
\begin{equation}
\label{eqn:imperfectfuse2}
	\vert \Psi \rangle = \left( \cos ( \theta_{b} ) \vert 0 \rangle + \sin (\theta_{b})\vert 1 \rangle  Z_{x}Z_{y}   \right)  P_{xy}(\gamma_{1})\vert \psi_{X} \rangle \vert \psi_{Y} \rangle ,
\end{equation}
which is the similar to Eqn.~(\ref{eqn:imperfectfuse}), except for the important addition of a partial fusion generated from previous attempts at merging, $P_{xy}(\gamma)$.  Again, we perform a rotation $M(\pm \theta_{b})$ on the central qubit, but this time allow for a choice in sign, the purpose of which will become evident soon.  This gives the state:
\begin{eqnarray}
\label{eqn:secondMergeResult1}
	\vert \Psi \rangle & = &  \sqrt{p_{s}(\theta_{b})}\vert 1 \rangle P_{xy}(\pm \pi/4) P_{xy}(\gamma_{1})\vert \psi_{X} \rangle \vert \psi_{Y} \rangle \\ \nonumber
	& & + \sqrt{1-p_{s}(\theta_{b})} \vert 0 \rangle  P_{xy}(\mp \gamma_{2}) P_{xy}(\gamma_{1}) \vert \psi_{X} \rangle \vert \psi_{Y} \rangle\; ,  
\end{eqnarray}
which can be simplified using the rules for combing partial fusions to give:
\begin{eqnarray}
\label{eqn:secondMergeResult2}
	\vert \Psi \rangle & = &  \sqrt{p_{m}(\theta_{b}, \gamma_{1})}\vert 1 \rangle P_{xy}(\pm \pi/4) \vert \psi_{X} \rangle \vert \psi_{Y} \rangle \\  \nonumber
	& & + \sqrt{1-p_{m}(\theta_{b}, \gamma_{1})} \vert 0 \rangle  P_{xy}(\gamma_{3}) \vert \psi_{X} \rangle \vert \psi_{Y} \rangle .
\end{eqnarray}
Here the value for $\gamma_{3}$ follows from Eqn.~(\ref{comPF2}), and the new amplitudes are products of the old amplitudes and the normalization constant, such that
\begin{equation}
	p_{m} (\theta_{b}, \gamma_{1})= p_{s}(\theta_{b}) N_{M}^{2}(\pm \pi/4 ,\gamma_{1})\; .
\end{equation}
This simplifies to
\begin{equation}
	p_{m} (\theta_{b}, \gamma_{1})= p_{s}(\theta_{b}) ( 1 \pm \sin ( 2 \gamma_{1})). 
\end{equation}
The second factor of this probability is due to the partial fusion, and by the correct choice of the sign $\pm$ can always be made greater than 1. This is achieved by matching the sign in the rotation $M(\pm \theta_{b})$ to the sign of $\gamma_{1}$.  Insight into the physical underpinnings of this sign matching can be gained from considering that partial fusions can be decomposed into linear sums of the odd and even parity projectors.  For a partial fusion $P_{xy}(\phi)$ we find the following: it is more even parity than odd when $-\pi/2 < \phi < 0$; and more odd parity than even when $0 < \phi < \pi/2$.  As for a rotation $M(\pm \theta)$, this will attempt to project onto the odd or even parity subspace for $+$ and $-$ respectively.  Since two graph state qubits initially have equal magnitude in the odd/even parity subspaces, a partial fusion will increase the magnitude in one particular subspace, and a measurement of parity is more likely to work for the dominant subspace.

As a closing remark on merging, notice that if a central vertex is not tilted it can be used deterministically to project the two qubits into a definite parity state.  If the two qubits are already partially fused then this alters the probability of an odd or even parity projection.  Some fine tunings of the strategy are described in the section~(\ref{sec:finetunings}).

\subsection{Bridging, and weighted graph edges}
\label{sec:bridge}

\begin{figure*}[t]
\centering
\includegraphics{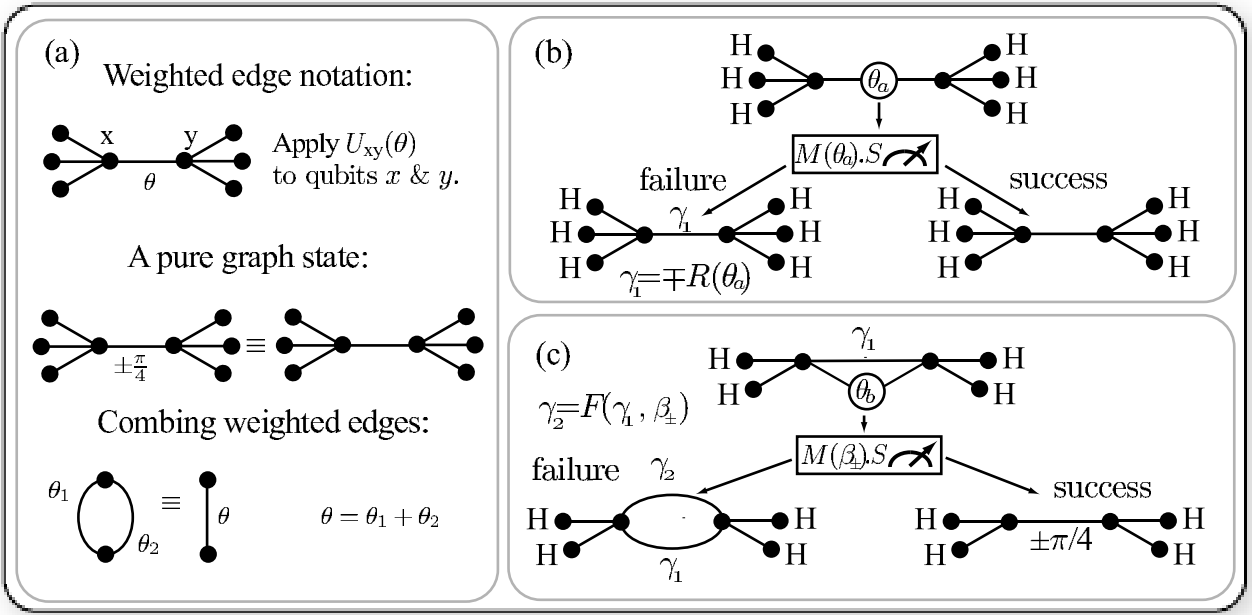}
\caption{The graphical notation for a weighted edge is defined and the procedure for bridging tilted vertices is shown.  (\textbf{a}) The symbol for a weighted edge, a graph edge labelled with an angle, is defined.  We also show how a maximally entangling partial fusion ($\theta=\pm \pi/4$) reduces to a pure graph state, and how multiple weighted edges combine into a single weighted edge. (\textbf{b}) The procedure for bridging tilted vertices when there is no pre-existing weighted. The rotation required before measuring the tilted vertex is $M(\theta_{a}).S$.  Upon failure, a weighted edge of angle $\mp R(\theta_{a})$ is created. (\textbf{c}) The procedure for bridging tilted vertices when there is a pre-existing weighted.  The required rotation prior to measuring the tilted vertex is $M(\beta_{\pm})$, where $\beta_{\pm}$ is defined in Eqn.~(\ref{BETA}). Upon failure a weighted edge is generated of angle $F(\gamma_{1}, \beta_{\pm})$, which is defined in Eqn.~(\ref{LF}).  Note that, to maximise the probability of success we must choose the correct sign in $\beta_{pm}$.}
\label{fig:WeightedEdges}
\end{figure*}

When performing a bridge operation on a pure graph state, as in Fig.~(\ref{fig:WeightedEdges}), it differs from a merge operation by an additional rotation $S$, where $S$ is a diagonal matrix with entries $( 1, i )$.  Again, we shall algebraically describe the bridge procedure in parallel with a graphical description given in Fig.~(\ref{fig:WeightedEdges}), and shall begin with a system resulting from the failed outcome of Fig.~(\ref{fig:Realign}b).  Initially the state is described by Eqn.~(\ref{eqn:imperfectfuse}) and after rotating the central qubit by $M(\pm \theta_{a}) \cdot S$ the state is:
\begin{eqnarray}
\label{Yprob}
	\vert \Psi \rangle & = &  \sqrt{p_{s}(\theta_{a})}\vert 1 \rangle U_{xy}(\pm \pi/4) \vert \psi_{X} \rangle \vert \psi_{Y} \rangle \\ \nonumber
	& & + \sqrt{1-p_{s}(\theta_{a})} \vert 0 \rangle U_{xy}(\gamma) \vert \psi_{X} \rangle \vert \psi_{Y} \rangle  \; .
\end{eqnarray}
Again, $\gamma$ is determined by the failure function such that $\gamma=\mp R(\theta_{a})$, and the new operator is:
\begin{equation}
U(\phi) = \cos (\phi) \unity + i \sin (\phi) Z_{x}Z_{y} ,
\end{equation}
which is always unitary.

As with the merging procedure there is a $p_{s}(\theta_{a})$ chance of successfully obtaining $\vert 1 \rangle$ in a measurement, and this generates a unitary $U_{xy}(\pm \pi/4)$ acting on $\vert \psi_{X} \rangle \vert \psi_{Y} \rangle$.  This unitary matrix is identical to a control-$Z$ operation with an additional $S_{x}S_{y}$ byproduct.  Again, failure establishes a partial amount of entanglement that can be recycled in later attempts at bridging.

As with the merging procedure, a failure on the first attempt is at least partially successful in that $U(\gamma)$ generates a control-$Z(-4 \gamma)$ gate with a byproduct $Z_{x}(-2 \gamma)Z_{y}(-2 \gamma)$. The notation $Z(\varphi)$ denotes the diagonal matrix with elements ($1,\exp^{i \varphi}$).  As before, this partial entanglement will improve the probability of success when a second attempt is made.  Furthermore, as with the merging procedure $M_{xy}(\theta) \cdot S$ and $M_{xy}(-\theta) \cdot S$ target two distinct but equally acceptable results, which generate either a control-$Z(\pi)$ or a control-$Z(-\pi)$. 

The above procedure is represented in Fig.~(\ref{fig:WeightedEdges}b), where the partial entanglement of a failed bridge is represented as a weighted edge, a solid line labelled with an angle.  From the definition of $U_{xy}(\theta)$, we can again derive a combination relation, $U_{xy}(\theta_{1})U_{xy}(\theta_{2})=U_{xy}(\theta_{1}+\theta_{2})$.  Therefore, unlike the partial fusion operator, $U_{xy}(\pm \pi/4)$ does not override previous failures.  Consequently, when later attempts are made at bridging it is necessary to target the correct amount of weighted edge.  If there is a preexisting edge of angle $\theta$ then a successful outcome is generated by $\pi/4 - \theta$ or $\pi/4 + \theta$.  As before, the sign choice will play an important role in recycling entanglement to boost to the probability of success. 

Taking a system with a weighted edge between $x$ and $y$ and then making, via nonideal double heralding, a tilted central vertex between $x$ and $y$, we have the state:
\begin{equation}
\label{eqn:imperfectfuse3}
	\vert \Psi \rangle = \left( \cos ( \theta_{b} ) \vert 0 \rangle  + \sin (\theta_{b})\vert 1 \rangle Z_{x}Z_{y} \right)  U_{xy}(\gamma_{1})\vert \psi_{X} \rangle \vert \psi_{Y} \rangle ,
\end{equation}
A weighted edge of magnitude $\pm \pi/4 - \gamma_{1}$ can be targeted by rotating the central qubit by $M(\beta_{\pm}) \cdot S$, where $\beta_{\pm}$ satisfies:
\begin{equation}
\label{BETA}
		\cos( \beta_{\pm} ) = N_{B} (\gamma_{1}, \theta_{b}) \cos (\theta_{b}) ( \pm \cos (\gamma_{1}) - \sin (\gamma_{1}) )\; .
\end{equation}
Here
\begin{equation}
		N_{B} (\gamma_{1}, \theta_{b}) = \left(1 \mp   \sin(2\gamma_{1}) \cos(2 \theta_{b}) \right)^{-\frac{1}{2}}
\end{equation}
plays a analogous role to $N_{M}$.  After the rotation, the state becomes:
\begin{eqnarray}
	\vert \Psi \rangle  =  \sqrt{p_{b} (\gamma_{1}, \theta_{b})} \vert 1 \rangle U(\pm \pi/4) \vert \psi_{X} \rangle \vert \psi_{Y} \rangle -  \\ 
	\sqrt{1-p_{b}(\gamma_{1}, \theta_{b})} \vert 0 \rangle U(\gamma_{1}+\gamma_{2}) \vert \psi_{X} \rangle \vert \psi_{Y} \rangle  \nonumber,
\end{eqnarray}
where $p_{b}(\gamma_{1}, \theta_{b})$ is the probability of success
\begin{equation}
	p_{b} (\gamma_{1},\theta_{b}) = N^{2}_{B} (\gamma_{1}, \theta_{b}) p_{s}(\theta_{b})\; .
\end{equation}
The angle $\gamma_{2}$ of the additional weighted edge equals $F(\gamma_{1} ,\beta_{\pm} )$:
\begin{equation}
\label{LF}
	\cos ( F (\gamma_{1}, \beta_{\pm}  ) ) = \frac{ \cos^{2}(\theta_{b}) \left( \pm \cos ( \gamma_{1}) - \sin (\gamma_{1}) \right) }  {\sqrt{N^{-2}_{B}(\gamma_{1}, \theta_{b})-p_{s}(\theta_{b} )}},
\end{equation}
where $F(\gamma_{1}, \beta_{\pm} )$ is a generalisation of the failure function $R(\phi)$, and it is simple to show that $R(\phi)=F(0, \phi )$.

The effect of the previous weighted edge modifies the success probability by a factor $N_{B}(\gamma_{1}, \theta_{b})$.  This factor can always be made greater than $1$ by the correct choice of the ``$\mp$'' sign, corresponding to a physical choice between in the targeted unitary $U(\pm \pi/4 - \gamma_{1})$.  The extent to which $N_{B}$ deviates from $1$ is dependent on $\sin(2 \gamma_{1})$, which is intuitive, as it vanishes for no pre-existing weighted edge ($\gamma_{1}=0$), and it is at a maximum for a pure graph edge ($\gamma_{1} = \pi/4$).  The effect of $\theta_{b}$ on $N_{B}$ is less intuitive, as it increases as the vertex approaches a qubit initialization in the $\vert 0 \rangle$ or $\vert 1 \rangle$ state.  Recall that a constructively defined graph state generates entanglement by control-$Z$ gates on initialized qubits, so any qubits in the $\vert 0 \rangle$ or $\vert 1 \rangle$ would not become entangled with the graph.  

Also, we find that the determinism of the ideal regime is regained when the central vertex is untilted  ($\theta_{b}=\pi/4$) even if there is a pre-existing weighted edge ($\gamma_{1} \neq 0$). Although the success probability does not become $1$, the failure outcome becomes the equally desirable alternative to the targeted operation, that is $U_{xy}(\mp \pi/4 - \gamma_{1})$ instead of the targeted $U_{xy}(\mp \pi/4 - \gamma_{1})$.  However, unlike merging in the ideal regime, $\gamma_{1}$ does not alter the probability of these outcomes, with each remaining equally likely. 

\subsection{overall strategies}
\label{sec:finetunings}

For both the merge and bridge procedure we have described the success probability when using a tilted central vertex of angle $\theta$.  However, there are alternative methods of generating this graph, which would result in a different values of $\theta$.  In either case, an attempt at double heralding has to be made, which will generate a tilted central vertex with a cherry, that is tilted by some amount $\theta_{a}$.  

For method (\textit{i}) the cherry is measured in the $Z$-basis, so that the tilted vertex can be used to merge or bridge with success probability:
\begin{equation}
 P_{i} =  N(\gamma, \theta_{a}) p_{s}(\theta_{a}),
\end{equation}
where $N$ is the appropriate factor, $N_{m}$ or $N_{b}$. 

For method (\textit{ii}) an attempt at realignment is made on the cherry, which if successful, guarantees success at merging or bridging, but if unsuccessful, makes changes the vertex tilting to $\theta_{\alpha}=-R(\theta_{a})$.  Hence, the overall probability of success is:
\begin{equation}\label{Pii}
 P_{ii} =  1 - [1 - N(\gamma, \theta_{\alpha}) p_{s}(\theta_{\alpha})][1 - p_{s}(\theta_{a})],
\end{equation}
which although benefiting from the additional opportunity at realignment, may suffer if $N(\gamma, \theta_{\alpha}) p_{s}(\theta_{\alpha})$ is substantially lower than $N(\gamma, \theta_{a}) p_{s}(\theta_{a})$.

When $\gamma \neq 0$ the best method can be determined by simple calculation.  On a first attempt at merging or bridging $\gamma=0$, and the best method is always the same.  If $\gamma=0$, then $N(0,\theta)=1$, and since $p_{s}(\theta_{\alpha})<p_{s}(\theta_{a})$, it follows that $P_{i}<P_{ii}$.  Note that on a first attempt $P_{ii}$ has at upper bound of $75 \%$, which is approached as $\theta_{a} \rightarrow \pi/4$, whereas on later attempts the upper bound can approach 1.

\section{Improvements in resource costs}
\label{ResOverhead}

The purpose of this paper has been to demonstrate that if we know what errors are caused in our entangling operation (due to the additional information provided by detector `click' times) then there are numerous strategies and procedures that we can deploy which allow us to construct a graph. This scheme is not designed to replace traditional error correction (as this will still be necessary), but rather to supplement it.  If we consider even a modest amount of cavity mismatch, the error rate for tilting errors is far beyond the current error correction thresholds.  Since these errors have to be prevented, the only alternative proposal that we are aware of is to post-select double heralding applications that meet some fidelity requirement.  We give some examples of how our proposal improves upon  a naive post-selection strategy.

\begin{figure}[t]
\centering
\includegraphics{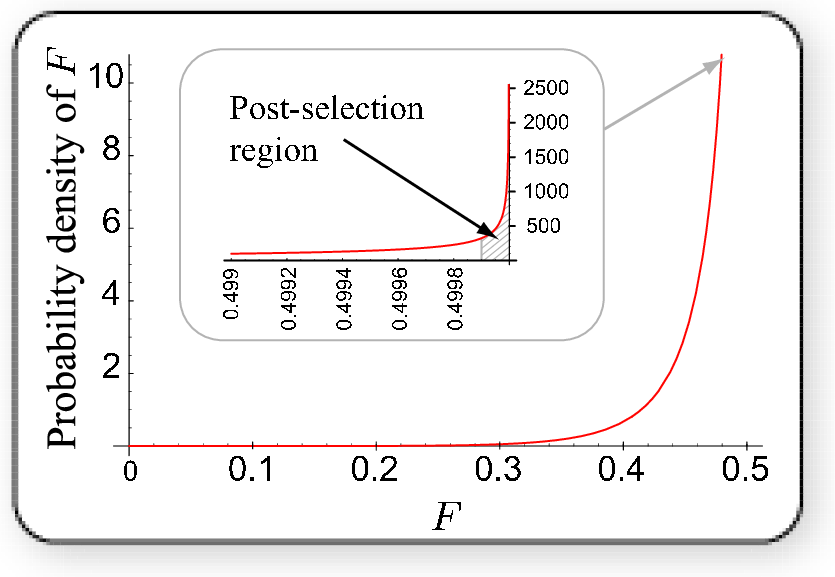}
\caption{The fidelity distribution for the example pair of cavities outlined in appendix \ref{sec:CavityExample}.  The inset is an enlargement of the region just below $0.5$, with the region of $0.5-10^{-4}<F<0.5$ highlighted, as this is the window of graph state that are post-selected as being within error correction tolerances.}
\label{fig:Improvements}
\end{figure}

As a measure of the two approaches we take the probability of successfully merging or bridging, on the first attempt.  The pair of cavities we use in our calculations are described in appendix~\ref{sec:CavityExample}, and their photon leakage rates are shown in Fig.~(\ref{Outline}c).  Taking the fidelity threshold such that $\frac{1}{2}-10^{-4}<F$, the post-selection scheme is constrained to the window in Fig.~(\ref{fig:Improvements}).  The area under this curve --- times the rate of photon loss --- is the probability of success, which we calculate to be $P($post-select$)=3.3\%$.  For our scheme, we can accept the same graph segments that the post-selection strategy accepts, and attempt to merge/bridge the lower fidelity graphs, hence $P($our-strategy$)=P($post-select$)+P($out-window$)$.  On the first attempt at merge/bridge Eqn.~(\ref{Pii}) simplifies to $3F^{2}$, and we calculate $P($outside-window$)$ by integrating $3F^{2} dF$ over the whole outside region, which gives $P($outside-window$)=35.7\%$.  Hence, $P($our-strategy$)=39.0\%$, and our strategy is an improvement by over an order of magnitude. 

Note that, an order of magnitude improvement in gate probability equates to far more than an order of magnitude saving in the resources (e.g. working memory) required by a quantum computer.  Indeed, although the scaling varies between different approaches to graph growth, it common to see resource costs that scale with $p^{- \log(n)}$ \cite{DR01a}, where $n$ is a measure of the computation size.  Hence, an order of magnitude improvement in gate success, is equates to an order $(0.1)^{\log(n)}$ reduction in resource costs.

\section{Conclusion}

In this paper a set of procedures have been proposed that allow distributed quantum computing to be achieved despite systematic errors that generate a tilting error in the graph vertices.  These errors have been shown to naturally arise when the double heralding scheme is used with photon sources that possess different inherent leakage rates.  Graph construction has been considered as consisting in three phases: \textit{phase 1} --- construction of a GHZ resource; \textit{phase-2} --- efficient merging of GHZ resources into larger GHZ states; \textit{phase-3} --- efficient bridging of GHZ states.  The proposed scheme contains modified protocols for each of these graph construction steps.  The modified protocols adapt to information from photon detection times so as to eliminate all known tilting error.  We have extended the graphical language of graphs states to include tilted vertices, weighted edges and partial fusions, so that an intuitive grasp of the errors can be gained.  Weighted graph edges have already found various applications throughout the field of quantum computation \cite{HHLB01a,HDERNB01a,DE01a}, and the authors suspect that tilted vertices and partial fusion may also prove to be useful concepts in other problems.

\section{Acknowledgements}

We would like to thank Dan Browne and Peter Rhode for useful comments on the manuscript.  This research is  part of the QIP IRC (GR/S82176/01). Simon Benjamin acknowledges support from the Royal Society.

\appendix

\section{Evaluating the integral to determine $E(F^{2})$}
\label{Integral}

The expression for $E(F^{2})$, the expectation value of the fidelity (minus $\frac{1}{2}$) squared, is a complicated function of $t_{1}$ and $t_{2}$ that does not directly integrate to a closed form.  However, we show that a solution can be found by restricting the expression to a region $R_{I}$ of $\theta_{a}$ and $\theta_{b}$.  Furthermore, a solution can be also be found for a second region $R_{J}$.  $R_{I}$ and $R_{J}$ overlap, and their union covers the entire region of interest, $0<\theta_{a}< \pi/2$ and $0<\theta_{b}< \pi/2$.  Hence, we can construct an analytic solution.  First, we express $F^{2}$ in the following form:
\begin{equation}
	E ( F^{2} ) = \frac{\Theta_{1}}{4} \int \left( \frac{U V}{U+V} \right)  \left( 1 + K \frac{V}{U+V}  \right)^{-1}   dt_{1} dt_{2} ,
\end{equation}
where $\Theta_{i}$ are defined in Eqn.~(\ref{eqn:BigTheta}), and:
\begin{equation}
 K = \left( \frac{\Theta_{1}}{\Theta_{2}} - 1 \right)
\end{equation}
and the functions $U$ and $V$ are:
\begin{eqnarray}
 U & = & P_{A}(t_{1})P_{B}(t_{2}), \\ \nonumber
 V & = & P_{B}(t_{1})P_{A}(t_{2}).
\end{eqnarray}
Since $U$ and $V$ are never negative, $V/(V+U)$ is always less than one.  Therefore, provided $K<1$ we will be able to perform a binomial expansion.  The restriction on $K$ gives us a region $R_{I}$, $\tan^{2}(\theta_{b})< 2 \tan^{2}(\theta_{a})$.  The expansion is:
\begin{equation}
	E( F^{2})_{R_{I}} = \sum_{n=1}^{\infty} (-1)^{n} \Theta_{1} \left( \frac{\Theta_{1}}{\Theta_{2}} - 1 \right)^{n} I_{n} ,
\end{equation}
where,
\begin{equation}
	I_{n} = \int_{0}^{\infty} \int_{0}^{\infty} U \left( \frac{V}{U+V} \right)  dt_{1} dt_{2}.
\end{equation}

\begin{figure}[t]
\centering
\includegraphics{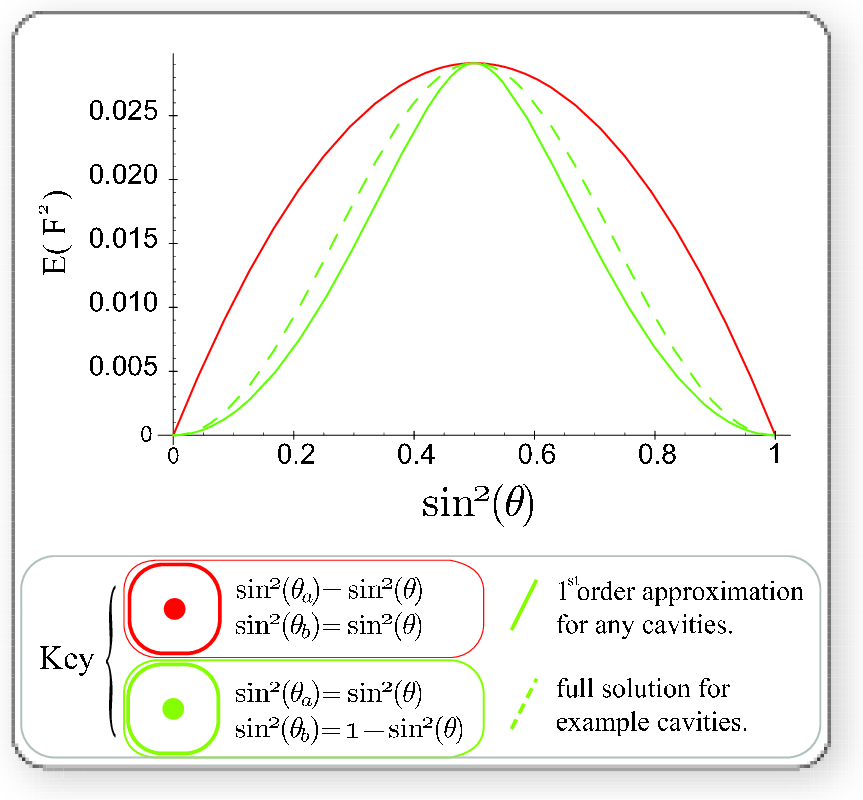}
\caption{The analytic $1^{st}$ order solution of $E(F^{2})$, compared to the full exact solution for a specific distribution.  The $1^{st}$ order solution is --- up to a factor --- independent of the photon leakage rates of the cavities.  These approximations are represented by solid lines, with the colour distinguishing cross-sections of $E(F^{2})$ along $\sin^{2}(\theta_{a})=\sin^{2}(\theta_{b})$  and $\sin^{2}(\theta_{a})=1- \sin^{2}(\theta_{b})$.  For the first cross-section, in red, the solution is exact even up to $0^{th}$ order.  For the second cross-section, in green, we compare the approximation with the a full solution, which is represented by a dashed line.  For the full solution we use the same cavity parameters as in Fig.~(\ref{Outline}c, \ref{fig:FidelitySQ}), which is specified in appendix \ref{sec:CavityExample}.}
\label{fig:FirstOrder}
\end{figure}

An expression for $E( F^{2})_{R_{I}}$ can be found by calculating $I_{n}$.  Although these may have to be performed numerically, this is preferable to numerically calculating $E(F^{2})$ at every point.  Also, we shall show that there is a recurrence relation between different $I_{n}$, so some terms can be found without any integration.  However, first we state the expansion for the second region $R_{J}$ ( $\tan^{2}(\theta_{a})< 2 \tan^{2}(\theta_{b})$):
\begin{equation}
	E( F^{2})_{R_{J}} = \sum_{n=1}^{\infty} (-1)^{n} \Theta_{2} \left( \frac{\Theta_{2}}{\Theta_{1}} - 1 \right)^{n} J_{n} ,
\end{equation}
where,
\begin{equation}
 J_{n} = \int_{0}^{\infty} \int_{0}^{\infty} V \left( \frac{U}{U+V} \right)  dt_{1} dt_{2}.
\end{equation}
Since the difference between $U$ and $V$ is only in the labelling of $t_{1}$ and $t_{2}$, we know $J_{n}$=$I_{n}$.  Furthermore:
\begin{eqnarray}
	J_{n} & = & \frac{UV}{U+V}\left( 1 - \frac{V}{U+V} \right)^{n} \\ \nonumber
	J_{n} & = & \sum_{k=0}^{n} (-1)^{k} \frac{n!}{(n-k)!k!} I_{n} 
\end{eqnarray}
Interestingly, for odd $n$, $I_{n}$ are fixed in terms of $I_{k}$ with even $k$. However, the even terms of $I_{n}$ are not constrained by this recurrence relation.  For example, $I_{1}=\frac{1}{2}I_{0}$, which means that (up to a constant factor $I_{0}$) we can expand up to first order without performing any numerical integration.  Hence, when $\sin^{2}(\theta_{a}) - \sin^{2}(\theta_{b}) \approx 0$,
\begin{equation} 
	E( F^{2}) \cong E(F^{2})_{aprx}=\left( \Theta_{L} - \frac{1}{2} \left( \frac{\Theta_{L}}{\Theta_{S}} -1 \right) \right) I_{0},
\end{equation}
where the change between $E( F^{2})_{R_{I}}$ and $E( F^{2})_{R_{J}}$, is achieved by using $\Theta_{L}$ as the largest of the pair $\Theta_{1}$ \& $\Theta_{2}$, and $\Theta_{S}$ as the smallest.  Two cross-sections of $E(F^{2})_{aprx}$ are shown in Fig.~(\ref{fig:FirstOrder}), and compared with an example full solution.  Both cross sections agree well with the exact solution, even far away from $\sin^{2}(\theta_{a}) - \sin^{2}(\theta_{b}) = 0$. The importance of this result is that it tells us about the shape of $E(F^{2})$ independently of the shape of $P_{A}(t)$ and $P_{B}(t)$.  In the main body of this paper we argue from a numerical solution of $E(F^{2})$ --- shown in Fig.~(\ref{fig:FidelitySQ}) --- that $\sin^{2}(\theta_{a})=\sin^{2}(\theta_{b})$ is preferable to $\sin^{2}{\theta_{a}}=1- \sin^{2}(\theta_{b})$.  The above result shows us that, at least near $\theta_{a} \approx \pi / 4$, this behaviour is not an artefact of the particular solution but a general property valid for any $P_{A}(t)$ and $P_{B}(t)$.  In fact, the only requirement is that $P_{i}$ are non-negative, so we can also accomodate photon loss and distributions that violate our Hamiltonian, Eqn.~($\ref{eqn:Ham}$).

\section{Example Cavity solutions}
\label{sec:CavityExample}

Here we outline solutions to the Jaynes-Cummings model and give an example solution that is used throughout the paper for expository purposes.  Solutions to Eqn.~(\ref{eqn:Ham}), involve two parameters, $g_{x}$ and $\kappa_{x}$.  In the strong coupling regime ($\kappa_{x} \ll 4g_{x}$),  the system tends to experience Rabi oscillations before the photon escapes from the system.  These oscillations will appear in the probability distribution for photon detection.  On the other hand, in the weak coupling regime ($\kappa_{x} \gg 4g_{x}$), the system experiences the quantum Zeno effect due to continual measurement of the cavity, and the relaxation time for the atom-cavity system increases with $\kappa_{x}$ as $\tau \sim \kappa_{x}/g_{x}^2$.  Arguably, this effect is an artefact of the model, as when the coupling decreases, it only decreases with respect to the cavity mode.  Therefore, in the weak coupling regime the system will lose photons faster than predicted, just into modes that are not being monitored with detectors.  However, this is not a major concern for our paper, as the primary purpose of invoking the Jaynes-Cummings Hamiltonian is to provide a means of modelling the collapse process.  The actual photon leakage rates, $P_{x}(t)$, will be determined experimentally.

Concerning the example solution: Rabi oscillations are undesirable for a system being used in double heralding; and the model is unsuitable for very weak coupling.  Between these regimes there exists a class of critically damped solutions --- for which $\kappa_{x} = 4g_{x}$ --- which have a convenient form that we take advantage of:
\begin{equation}
	C_x(t,g_x ) = P_{x}(t,g_x, 4g_x) = 4 g_{x}^{3} t^2 \exp^{-2 g_{x} t} \Theta (t),
\end{equation}
\noindent where $\Theta (t)$ is the Heavyside function
\begin{eqnarray}
	\Theta (t) =	 \begin{array}{lll}
0  					 &   &  t \leq 0       \\
1            &   &  t > 0   . \\ 
 \end{array} 
\end{eqnarray}
The example pair of cavities used throughout this paper, are parametrized by $g_{A}=10$ and $g_{A}=12.5$.  Note, that all conclusions of this paper are independent of the particular form of the photon leakage rate, and that this example is solely used to guide the reader through the paper. 


\end{document}